\documentclass[epjCONF]{svjour}
\usepackage{graphics}
\usepackage[varg]{txfonts} 
\usepackage[latin1]{inputenc}
\session-title{MESON2012 - 12th International Workshop on Meson Production, Properties and Interaction} 
\begin{document}
\title{Study of the $\eta \to e^+e^-\gamma$ Decay Using WASA-at-COSY Detector System}
\author{Ma{\l}gorzata Hodana\inst{1}\inst{2}\fnmsep\thanks{\email{m.hodana@gmail.com}} \and Pawe{\l} Moskal\inst{1}\inst{2} for the WASA-at-COSY Collaboration }
\institute{Institute of Physics, Jagiellonian University, PL-30059 Cracow, Poland \and Institut f\"{u}r Kernphysik and J\"{u}lich Center for Hadron Physics, D-52425 J\"{u}lich, Germany }
\abstract{
At the turn of October and November 2008, the WASA-at-COSY collaboration performed an experiment
to collect data on the $\eta$ meson decays in $pd \rightarrow {}^{3}He\, \eta$ reactions. This report presents results of a 
study of the $\eta \rightarrow e^+ e^- \gamma $ conversion decay. 
The desired feature of this decay is that the final state $e^+ e^-$ pairs come from 
the conversion of a virtual $\gamma$ quantum and, therefore, they constitute a rich source of knowledge 
about the electromagnetic structure of decaying meson. 
This is a very important characteristic when it comes to study of a short-lived neutral particle
like the $\eta$ meson, because the classical methods of particle scattering are not applicable here. 
The analysis performed on one third of the p-d data sample collected using the WASA-at-COSY detector, lead to the extraction 
of the $\eta$ transition form factor as a function of the $e^+ e^-$ mass and to the calculation 
of the slope parameter, related to the charge radius of the $\eta$ meson.
} 
\maketitle
\section{Introduction}
\label{intro}
For the $\eta$ meson is a short-lived, neutral particle, it is not possible to investigate its structure via the classical method of particle scattering. To learn about its quark wave function, one studies those decay processes of this meson, in which a pair of photons is produced, at least one of them being virtual. The virtual photons have a non-zero mass and convert into lepton-antilepton pairs. The squared four-momentum transferred by the virtual photon corresponds to the squared invariant mass of the created lepton-antilepton pair. Therefore, information about the quarks' spatial distribution inside the meson can be achieved from the lepton-antilepton invariant mass distributions by comparison of empirical results with predictions, based on the assumption that the meson is a point-like particle. The last can be obtained from the theory of Quantum Electrodynamics. The deviation from the expected behavior in the leptonic mass spectrum expose the inner structure of the meson. This deviation is characterized by a form factor. 
It is currently not possible to precisely predict the dependence of the form factor on the four-momentum transferred by the virtual photon in the framework of Quantum Chromodynamics theory. Therefore, to conduct calculations, assumptions about the dynamics of the investigated decay are needed.

The knowledge of the form factors is also important in studies of the muon anomalous magnetic moment, ${a_{\mu} = (g_{\mu}-2)/2}$, which is the most precise test of the Standard Model and, as well, may be an excellent probe of new physics. The theoretical error of calculation of $a_{\mu}$ is dominated by hadronic corrections and therefore limited by the accuracy of their determination. 
Especially, the hadronic light-by-light scattering contribution to $a_{\mu}$ includes two meson-photon-photon vertices and therefore also depends on the form factors \cite{Bijnens:1999jp}.
At present, the discrepancy between the  $a_{\mu}$ prediction based on the Standard Model and its experimental value \cite{Bennett:2006fi} is equal to ${(28.7 \pm 8.0) \cdot 10^{-10}}$  ($3.6\sigma$) \cite{Davier:2010nc}. 
\section{Experiment and Results}
\label{sec:1}
The experiment was performed at the end of 2008 by the WASA-at-COSY collaboration. The COSY facility provided a proton beam of momentum $1.7$~GeV/c which has been used to produce $\eta$ mesons by collisions with deuteron target. 

All particles have been measured explicitly, taking advantage of the large
acceptance of the WASA $4\pi$ detector. Helium ions were registered using the
forward part of the detector while the central detector's part was used to
measure leptons and photons (in the calorimeter). 
Particles' charge was determined based on the magnetic field provided by the
superconducting solenoid surrounding the drift chamber of the Central
Detector. A detailed description of the detector system can be found in \cite{Adam:2004ch}.

The analysis performed so far allowed
for the extraction of $525 \pm 26$ events of
the single Dalitz decay, up to the mass
of $e^+ e^-$ pairs equal to $0.425~\mathrm{GeV/c^2}$ \cite{hod_phd}.

Applied restrictions allowed to suppress the background from other $\eta$ decays to a negligible level and the multipion 
background was subtracted from the signal, based on missing mass distributions evaluated for each $M_{e^+ e^-}$ separately.
This allowed for the extraction of the $\eta$ transition form factor as a function of the $M_{e^+ e^-}$ mass 
and, therefore, for the calculation of the slope parameter of the structure
function which provides the knowledge of the meson-photon transition region.

\begin{figure}
\centering
\resizebox{0.65\columnwidth}{!}{%
	\includegraphics{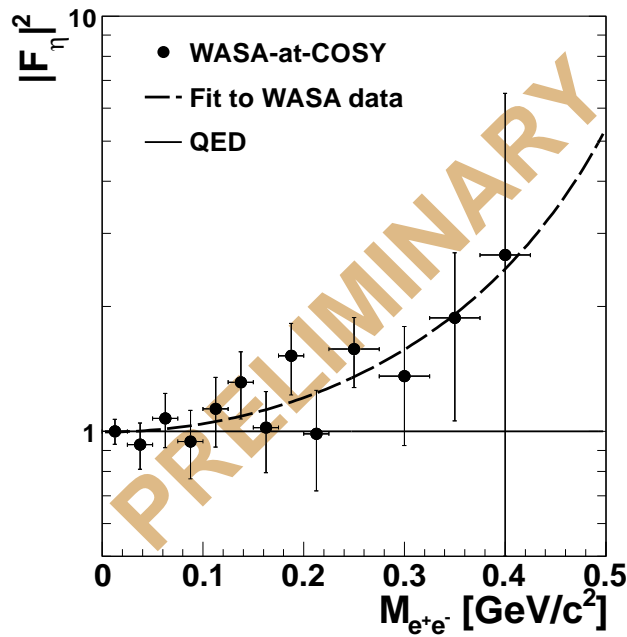}}
	\caption{Experimental spectrum of the squared transition form factor, $|F_\eta|^{2}$, as a function of the $M_{e^+ e^-}$. The dashed line is the result of the fit of the single-pole formula to experimental points. The solid line is the QED model assumption of a point-like meson. }
	\label{fig:FF}
\end{figure}

The resulting distribution of the transition form factor squared, $|F_\eta|^{2}$, as a function of the invariant mass of the leptons pairs, $M_{e^+ e^-}$ is shown in Fig. \ref{fig:FF}. This structure function of the $\eta$ meson corresponds to the ratio of the
experimentally obtained distribution of the $e^+ e^-$ mass to the model prediction in
which the transition form factor is constant in the four-momentum transfer and equals one.

The obtained value of the slope parameter of the structure function of the $\eta$ meson is 
\begin{equation}
b_P = \frac{dF_{\eta}}{dq^2}|_{q^2\simeq0} = \Lambda^{-2} \nonumber 
    = (2.27 \pm 0.73_\mathrm{stat.} \pm 0.46_\mathrm{sys.})\,\mathrm{GeV}^{-2},
\end{equation}
where the estimated systematical uncertainty should be treated as an upper limit only.

This preliminary result is consistent with the one obtained by CB/TAPS collaboration, $b_P^{\mathrm{CB/TAPS}} = (1.92 \pm 0.35_\mathrm{stat.} \pm 0.13_\mathrm{sys.})$~$\mathrm{GeV}^{-2}$ \cite{Berghauser:2011zz} and with the one obtained using $\eta \to \mu^{+}\mu^{-}(\gamma)$ decays studied in the heavy ion experiment NA60 in which photons were not registered, $b_P^{\mathrm{NA60}} = (1.95 \pm 0.17_\mathrm{stat.} \pm 0.05_\mathrm{sys.}$)~$\mathrm{GeV}^{-2}$ \cite{Arnaldi:2009wb}. 

The three results mentioned above, enable to estimate the charge radius of the $\eta$ meson,
\begin{equation}
<r_{\eta}^{2}>^{1/2} = \sqrt{6\cdot b_P} \nonumber = 0.68 \pm 0.02 ~{\mathrm{fm}}.
\end{equation}
This value is in disagreement, at the level of two standard deviations, with the one calculated using the Vector Meson Dominance model, where ${<r_{\eta}^{2}>^{1/2} = 0.64}$~fm \cite{Ametller:1991jv}. It is also interesting to notice that the obtained radius of the $\eta$ meson is smaller than the radius of charged pion of
$<r_{\pi}^{2}>^{1/2} = (0.74 \pm 0.03)$~fm \cite{Liesenfeld:1999mv}.

\section{Acknowledgments}
\label{sec:2}
We acknowledge support by the  FFE grants of the Research Center Juelich and 
by the Polish Ministry of Science 
and Higher Education under grants No. 0320/B/H03/2011/40 and UMO-2011/01/B/STL/00431.

\end{document}